\documentclass[useAMS,usenatbib]{mn2e}

\usepackage[dvips]{graphicx}
\usepackage{epsfig,natbib}
\usepackage{mathptm}

\def\pmb#1{\setbox0=\hbox{#1}%
  \kern-0.02em\copy0\kern-\wd0
  \kern+0.04em\copy0\kern-\wd0
  \kern-0.02em\copy0\kern-\wd0
  \raise+0.02em\copy0\kern-\wd0
  \raise-0.02em\box0}
\def\bell{\pmb{$\ell$}}

\def\reff@jnl#1{{\rm#1\/}}
\def\aj{\reff@jnl{AJ}}                 
\def\araa{\reff@jnl{ARA\&A}}           
\def\apj{\reff@jnl{ApJ}}               
\def\apjl{\reff@jnl{ApJ}}              
\def\apjs{\reff@jnl{ApJS}}             
\def\ao{\reff@jnl{Appl.Optics}}        
\def\apss{\reff@jnl{Ap\&SS}}           
\def\aap{\reff@jnl{A\&A}}              
\def\aapr{\reff@jnl{A\&A~Rev.}}        
\def\aaps{\reff@jnl{A\&AS}}            
\def\azh{\reff@jnl{AZh}}               
\def\baas{\reff@jnl{BAAS}}             
\def\jrasc{\reff@jnl{JRASC}}           
\def\memras{\reff@jnl{MmRAS}}          
\def\mnras{\reff@jnl{MNRAS}}           
\def\pra{\reff@jnl{Phys.Rev.A}}        
\def\prb{\reff@jnl{Phys.Rev.B}}        
\def\prc{\reff@jnl{Phys.Rev.C}}        
\def\prd{\reff@jnl{Phys.Rev.D}}        
\def\prl{\reff@jnl{Phys.Rev.Lett}}     
\def\pasp{\reff@jnl{PASP}}             
\def\pasj{\reff@jnl{PASJ}}             
\def\qjras{\reff@jnl{QJRAS}}           
\def\skytel{\reff@jnl{S\&T}}           
\def\solphys{\reff@jnl{Solar~Phys.}}   
\def\sovast{\reff@jnl{Soviet~Ast.}}    
\def\ssr{\reff@jnl{Space~Sci.Rev.}}    
\def\zap{\reff@jnl{ZAp}}               
\def\nat{\reff@jnl{Nature}}            
\def\be{\begin{equation}}
\def\ee{\end{equation}}
\def\ber{\begin{eqnarray}}
\def\eer{\end{eqnarray}}

\title{Simulation of non-Gaussian CMB maps}
\author[Graca Rocha et al.]
{Graca Rocha$^{1,2,3}$, M.P.~Hobson$^1$, Sarah Smith$^1$, Pedro
  Ferreira$^2$ and Anthony Challinor$^1$\\
$^1$Astrophysics Group, Cavendish Laboratory, Magingley Road,
Cambridge CB3 0HE, UK\\
$^2$Astrophysics, Denys Wilkinson Building, Keble Road, Oxford OX1 3RH, UK\\
$^3$Centro de Astrof\'{\i}sica da Universidade do Porto, R. das
Estrelas s/n, 4150-762 Porto, Portugal}

\date{Accepted ---. Received ---; in original form \today}

\pagerange{\pageref{firstpage}--\pageref{lastpage}} \pubyear{2004}

\begin{document}

\label{firstpage}

\maketitle

\begin{abstract}
\noindent A simple method is presented for the rapid simulation of
statistically-isotropic non-Gaussian maps of CMB temperature
fluctuations with a given power spectrum and analytically-calculable
bispectrum and higher-order polyspectra.  The $n$th-order correlators
of the pixel values may also be calculated analytically. 
The cumulants of the simulated map may be used to obtain
an expression for the probability density function of the pixel
temperatures.  The statistical properties of the simulated map are
determined by the univariate non-Gaussian distribution from which
pixel values are drawn independently in the first stage of the
simulation process. We illustrate the method using a non-Gaussian
distribution derived from the wavefunctions of the harmonic
oscillator. The basic simulation method is easily extended to produce
non-Gaussian maps with a given power spectrum and diagonal bispectrum.

\end{abstract}



\section{Introduction}
\label{sec:intro}

The study of non-Gaussianity of Cosmic Microwave Background (CMB) fluctuations is of major
importance in understanding the processes responsible for generating
the fluctuations and in assessing the contribution of foreground
astrophysical processes and instrumental effects to observations of
the CMB. 

Single-field inflationary scenarios predict, in general, that CMB
fluctuations are very nearly Gaussian (see e.g.~\citealt{bartolo04}
for a recent review), if one assumes that the
sub-Hubble-scale quantum fluctuations start off in the ground state
\citep*{joao,martin,gangui}. In such models, non-Gaussianity produced
during inflation arises
predominantly from the non-linear nature of gravitational interactions
rather than from self-interaction of the fluctuations of the inflaton
field. Typically the level of non-Gaussianity is suppressed by the first-order
slow-roll parameters~\citep{Acquaviva,Maldacena}. Subsequent non-linear
processing of the primordial fluctuations to second-order in perturbation
theory has been shown to amplify the tiny primordial non-Gaussianity to
a level on the last-scattering surface that may be detectable with future
CMB surveys~\citep{Bartolo}. Furthermore, second-order radiative transfer
effects, such as gravitational lensing of the CMB, should produce
a detectable level of non-Gaussianity in the CMB~\citep{zaldarriaga00}.
Larger levels of non-Gaussianity can be produced in inflation models
with multiple scalar fields. Examples include the curvaton
(e.g.~\citealt{lyth02}) and the inhomogeneous
reheating~(e.g.~\citealt*{dvali04}) scenarios.
Finally models which include topological defects also produce significantly
non-Gaussian fluctuations~\citep*{gangui01}.
It is also the case that inevitable
contaminants, such as discrete radio sources, Galactic emission and
systematic instrumental effects leave non-Gaussian signatures on
CMB maps. Thus non-Gaussianity tests are of fundamental importance
both for probing inflation physics and for isolating systematic effects. 

In order to investigate one's ability to detect and recover
non-Gaussian signals, it is useful to generate non-Gaussian maps with
known statistical properties, to which putative analysis methods may
be applied.  In particular, in many applications, it is desirable that
the simulated non-Gaussian map is statistically isotropic, with a
prescribed power spectrum (or 2-point correlation function).  The
generation of such non-Gaussian CMB maps is, however, a surprisingly
difficult task (see, for example, \citealt{vio01,vio02}).  Moreover, it
is often useful for the non-Gaussian map also to have a known (or
prescribed) bispectrum and one-point marginal probability density
function.  Several different techniques have been proposed to address
various subsets of these requirements \citep{contaldi, Martinez,
komatsu, liguori}, but each carries a considerable computational
cost. The aim of this paper is to present a simple, fast technique for
simulating statistically-isotropic non-Gaussian CMB maps with a
prescribed power spectrum, for which one can calculate analytically
the bispectrum, higher-order polyspectra, $n$th order pixel
correlators and the one-dimensional marginalised distribution (in
terms of its cumulants). Moreover, our basic simulation method is
easily extended to produce non-Gaussian maps with a given power
spectrum and diagonal bispectrum.

The problem of simulating non-Gaussian CMB maps can be formalised as
follows. A real, random scalar field $T(\mathbfit{x})$ can be defined
as a collection of random variables, one at each point
$\mathbfit{x}=(x_1,x_2,\ldots,x_n)$ in the $n$-dimensional space
(clearly $n=2$ for CMB maps on the sphere or flat-patches of sky).
Thus, for each position $\mathbfit{x}$,
$T(\mathbfit{x})\equiv t$, where $t$ is a scalar random variable
with a one-dimensional (marginal) probability density function (PDF)
$f_T(t)$. For a random field that is statistically homogeneous,
$f_T(t)$ is the same at all points in the space.  The main difficulty
in the numerical simulation of a generic random field is that, in
general, given two arbitrary positions $\mathbfit{x}_1$ and
$\mathbfit{x}_2$, the quantities $T(\mathbfit{x}_1)$ and
$T(\mathbfit{x}_2)$ are not independent. In particular, it is often
desirable for $T(\mathbfit{x})$ to have a prescribed 2-point
covariance function
\[
\xi_T(\mathbfit{x}_1,\mathbfit{x}_2) = \langle T(\mathbfit{x}_1)
T(\mathbfit{x}_2)\rangle.
\]
In the case of a statistically homogeneous random field in Euclidean
space, the covariance function depends only on $\btau \equiv
\mathbfit{x}_1-\mathbfit{x}_2$. If the field is also isotropic then
the dependence is only on $\tau = |\btau|$.

As discussed by \citealt{vio01,vio02}, most methods for simulating
non-Gaussian maps with a prescribed 2-point covariance function [and a
prescribed marginal PDF $f_T(t)$] are based on first generating a
zero-mean, unit-variance Gaussian random field $G(\mathbfit{x})$, with
an appropriate covariance structure $\xi_G(\tau)$. One then performs
the mapping transformation $G(\mathbfit{x}) \to T(\mathbfit{x})$
according to
\[
T(\mathbfit{x}) = h[G(\mathbfit{x})],
\]
where $h$ represents an appropriate function. The usefulness of this
approach derives from the fact that there exist explicit analytical
(but complicated) formulae for the marginal PDF, $f_T(t)$, and the
covariance function, $\xi_T(\tau)$, of the transformed field in terms
of the covariance function, $\xi_G(\tau)$, of the original Gaussian
field and the mapping function $h$. In particular, we note that the
formula for $\xi_T(\tau)$ takes the form of a double integral of a
function depending on both $\xi_G(\tau)$ and $h$. There exist only a
few mapping functions $h$ for which $f_T(t)$ and $\xi_T(\tau)$ may be
calculated analytically. A still smaller subset of these cases allows
the resulting expressions to be inverted analytically to obtain the
required functions $\xi_G(\tau)$ and $h$ to be used in simulating the
non-Gaussian map. In general, one has to resort to numerical methods
to invert the general formulae for $f_T(t)$ and $\xi_T(\tau)$ and this
can be computationally very costly.

As mentioned above, it is often desirable for the simulated
non-Gaussian map also to have a known (or prescribed) bispectrum.  The
method outlined above has not been extended to this case, and any such
generalisation is likely to be extremely computationally demanding.
An alternative method for generating non-Gaussian maps with a
prescribed power spectrum and bispectrum has been suggested by
\citet{contaldi}, although, in general, the marginal distribution
$f_T(t)$ of the resulting map cannot be obtained analytically. The
method is based on choosing some one-dimensional non-Gaussian PDF,
from which the real and imaginary parts of (some subset of) the
spherical harmonic coefficients $a_{\ell m}$ of the map are drawn
independently (the remaining coefficients being drawn from a Gaussian
PDF). However, statistical isotropy imposes `selection rules' upon
correlators of the $a_{\ell m}$ coefficients. For example, the
third-order correlators must satisfy
\[
\langle a_{\ell_1 m_1} a_{\ell_2 m_2} a_{\ell_3 m_3} \rangle
= \left(
\begin{array}{ccc}
\ell_1 & \ell_2 & \ell_3 \\
m_1 & m_2 & m_3 
\end{array}
\right) B_{\ell_1 \ell_2 \ell_3},
\]
where $(\cdots)$ denotes the Wigner $3j$ symbol and $B_{\ell_1 \ell_2
\ell_3}$ are the bispectrum coefficients.  Hence the map corresponding
to the drawn $a_{\ell m}$ values is not only non-Gaussian but also
anisotropic since all its third-order correlators are zero, except for
a subset of the form $\langle a_{\ell m}^3 \rangle$ for those $a_{\ell
m}$ drawn from the non-Gaussian PDF.  It is therefore necessary first
to produce an ensemble of non-Gaussian, anisotropic maps and then
create an isotropic ensemble by applying a random rotation to each
realisation. \citet{contaldi} show that these random rotations produce
the necessary correlations between the $a_{\ell m}$ coefficients to
ensure isotropy of the ensemble, but the method clearly requires
significant computation.

In this paper, we discuss a very simple and computationally fast
method for simulating non-Gaussian maps that are, by construction,
statistically isotropic, and for which numerous statistical properties
may be calculated analytically. In particular, it is possible to
produce a map with a prescribed power spectrum for which one can
obtain simple analytical expressions for the bispectrum and
higher-order polyspectra, and $n$th-order correlators of the pixel
values. One may also calculate the cumulants of the map, which may be
used to obtain the one-dimensional marginalised distribution.  A
simple extension of the method allows for the simulation of maps with
a prescribed power spectrum and diagonal bispectrum.  In
Section~\ref{sec:sim_method}, we discuss our method for simulating
non-Gaussian maps, the statistical properties of which are presented
in Section~\ref{sec:stat_props}.  
In Section~\ref{sec:extsim} we extend the method to allow simulation of maps with 
prescribed power spectrum and diagonal bispectrum.
Finally, our conclusions are
presented in Section~\ref{sec:conc}.


\section{Simulation method}
\label{sec:sim_method}

Since our goal is the simulation of CMB maps for use in later
analysis, it is convenient at the outset to divide the celestial
sphere into pixels labelled by $p = 1,2,\ldots,N_{\rm pix}$. For
simplicity, we also assume an equal-area pixelisation, so that each
pixel subtends the same solid angle $\Omega_{\rm pix} =4\pi/N_{\rm
pix}$.  Examples of such pixelisation schemes are {\sc
Healp}ix\footnote{http://www.eso.org/science/healpix/}~(\citealt{healpix})
and {\sc Igloo}~(\citealt{igloo}). The distribution of pixel centres
across the sphere is unimportant.

\begin{figure}
\begin{center}
\psfig{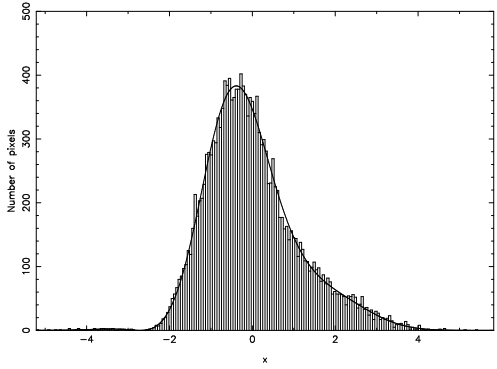}
\end{center}
\caption{The non-Gaussian PDF used in the simulations (solid line), 
which has the form given in equation~(\ref{like}) with 
$\alpha_{3}=0.2$ and $\sigma_{0}=1$, and a
set of samples drawn from the PDF (histogram).}
\label{fig:sho} 
\end{figure}

Our simulation method begins by drawing each pixel value $s_p\equiv
S(\mathbfit{x}_p) = S(\theta_p,\phi_p)$ independently from the same
one-dimensional non-Gaussian PDF $f_S(s)$.  The precise PDF used is
unimportant, but for the purposes of illustration, we adopt here a PDF
derived from the Hilbert space of a linear harmonic oscillator, as
developed by \citet{rocha}. This class of PDF is summarised in
Appendix A.  In particular, we assume the PDF illustrated in
Fig.~\ref{fig:sho}, which is chosen for convenience to have a mean of
zero.

The resulting map $S(\mathbfit{x})$ will, by construction, be
statistically isotropic to within the errors introduced by the
pixelisation scheme.  In fact, it is a realisation of isotropic
non-Gaussian white noise, with a variance given by the second
(central) moment $\mu_2$ of the PDF $f_S(s)$. We will assume
throughout that the mean $\mu_1$ of the generating non-Gaussian PDF is
zero, so that moments and central moments coincide.  According to the
`cumulant expansion theorem'~\citep{Ma:1985} the cumulants, or
connected moments, of the multivariate distribution of pixel values
are related to the logarithm of the moment-generating function,
\begin{equation}
M(\mathbfit{k}) \equiv \langle \rmn{e}^{\rmn{i}\mathbfit{s}\cdot \mathbfit{k}}
\rangle ,
\end{equation}
where $\mathbfit{s} \equiv (s_1,\dots,s_{N_\rmn{pix}})$ is the vector
of pixel values and $\mathbfit{k} \equiv
(k_1,\dots,k_{N_\rmn{pix}})$. The connected moments are given in terms
of the derivatives of $\ln M(\mathbfit{k})$ by
\begin{equation}
\left< s_{p_1} \cdots s_{p_n} \right>_c = (-\rmn{i})^n \frac{\partial^n}{\partial
k_{p_1}  \cdots \partial k_{p_n}}
\ln M(\mathbfit{k}) \Big|_{\mathbfit{k}=0}.
\label{eq:cet}
\end{equation}
Given that the pixel values $s_p$ are independent random variables we
have $\ln M(\mathbfit{k})= \sum_{p} \ln M_S(k_p)$, where
\begin{eqnarray}
\ln M_S(k) &\equiv& 
\ln \langle \rmn{e}^{\rmn{i}sk} \rangle = \ln \int_{-\infty}^{\infty}
f_S(s) \rmn{e}^{\rmn{i}sk}\, \rmn{d}s \nonumber \\
&=& \sum_{n=1}^\infty \frac{(\rmn{i}k)^n}{n!} \kappa_n
\end{eqnarray}
is the logarithm of the moment-generating function (the
cumulant-generating function) for the non-Gaussian PDF $f_S(s)$ whose
cumulants are the $\kappa_n$. Substituting in equation~(\ref{eq:cet}),
we find 
\be \left< s_{p_1} \cdots s_{p_n} \right>_c = \delta _{{p_1}
\cdots {p_n}} (- \rmn{i})^n \frac{\rmn{d}^n}{\rmn{d}k^n}\ln M_S(k)
\Big|_{k=0} = \delta_{{p_1} \cdots {p_n}} \kappa_n.
\label{eqn:cuma}
\ee 
The symbol $\delta_{p_1 \cdots p_n}$ equals one if all the $n$
pixels are the same and vanishes otherwise.  If required, the pixel
correlations can always be expanded in their connected
parts~\citep{Ma:1985}. In particular, since $\langle s_p \rangle = 0$
for each pixel, one finds that, for example,
\begin{eqnarray*}
\langle s_{p_1}s_{p_2} \rangle & = & \langle s_{p_1}s_{p_2} \rangle_c \\
\langle s_{p_1}s_{p_2}s_{p_3} \rangle & = & 
\langle s_{p_1}s_{p_2}s_{p_3} \rangle_c \\
\langle s_{p_1}s_{p_2}s_{p_3}s_{p_4} \rangle 
& = & \langle s_{p_1}s_{p_2}s_{p_3}s_{p_4} \rangle_c +
\langle s_{p_1}s_{p_2} \rangle_c \langle s_{p_3}s_{p_4} \rangle_c \\
& & + \langle s_{p_1}s_{p_3} \rangle_c \langle s_{p_2}s_{p_4} \rangle_c 
+\langle s_{p_1}s_{p_4} \rangle_c \langle s_{p_2}s_{p_3} \rangle_c.
\end{eqnarray*}

The next step in the simulation procedure is to transform the map
$S(\mathbfit{x})$ into spherical-harmonic space (using, for example,
the {\tt map2alm} routine from the {\sc Healp}ix package) to obtain
the coefficients
\be
a_{\ell m} = \sum_{p=1}^{N_{\rm pix}} Y^*_{\ell m} (\mathbfit{x}_p) s_p
\Omega_{\rm pix}
\approx \int_{4\pi} Y^*_{\ell m}(\theta,\phi)
S(\theta,\phi)\,\rmn{d}\Omega.
\label{eqn:almdef}
\ee
Using equation (\ref{eqn:cuma}) and the (approximate) orthogonality of
the (pixelised) spherical harmonics, we quickly find that the
second-order correlator of the harmonic coefficients is given by
\begin{equation}
\langle a_{\ell m} a^*_{\ell' m'} \rangle
= \mu_2 \Omega_{\rm pix} \delta_{\ell\ell'}\delta_{mm'},
\end{equation}
where $\mu_2 = \kappa_2$ since $f_S(s)$ has vanishing mean.  In order
to obtain a final non-Gaussian map with a particular prescribed
ensemble-average power spectrum, $C_\ell$, one then rescales the
harmonic coefficients to obtain
\be
\bar{a}_{\ell m} = a_{\ell m} \sqrt{\frac{C_\ell}{\mu_2\Omega_{\rm pix}}},
\label{eqn:abar}
\ee
such that $\langle \bar{a}_{\ell m} \bar{a}^*_{\ell' m'} \rangle =
C_\ell \delta_{\ell\ell'}\delta_{mm'}$.  We note that the effect of a
spatially-invariant, circularly-symmetric observing beam on the final
map is trivially included by letting $C_\ell \to C_\ell B_\ell^2$,
where $B_l$ are the coefficients of the beam in a Legendre expansion.
Finally, the harmonic coefficients $\bar{a}_{\ell m}$ are inverse
spherical harmonic transformed (using, for example, the {\tt alm2map}
routine from the {\sc Healp}ix package) to obtain the final map
\be
t_p \equiv T(\mathbfit{x}_p) = \sum_{\ell, m}\,
\bar{a}_{\ell m} Y_{\ell m}(\mathbfit{x}_p),
\label{eqn:tpdef}
\ee
where the double summation extends from $\ell=0$ to $\ell_{\rm max}$
and $m=-\ell$ to $\ell$. The equivalent flat-sky approximation for
small patches is discussed in Appendix~\ref{sec:flat_sky}.

In Fig.~\ref{fig:ngsimfull} we plot a realisation of a non-Gaussian
all-sky CMB map generated as described above, using the non-Gaussian
PDF plotted in Fig.~\ref{fig:sho} with a prescribed ensemble-average
power spectrum, $C_\ell$. The map was produced using the {\sc Healp}ix
pixelisation scheme with the $N_{\rmn{side}}$ parameter set to 512,
which corresponds to $N_{\rmn{pix}}=12 N_{\rmn{side}}^2 \approx
3\times 10^6$ equal-area pixels. For comparison, in
Fig.~\ref{fig:gsimfull}, we plot a realisation of a Gaussian CMB map
with the same power spectrum $C_\ell$, using the same pixelisation.
\begin{figure}
\begin{center}
\includegraphics[height= 0.35\textheight,angle = 90]{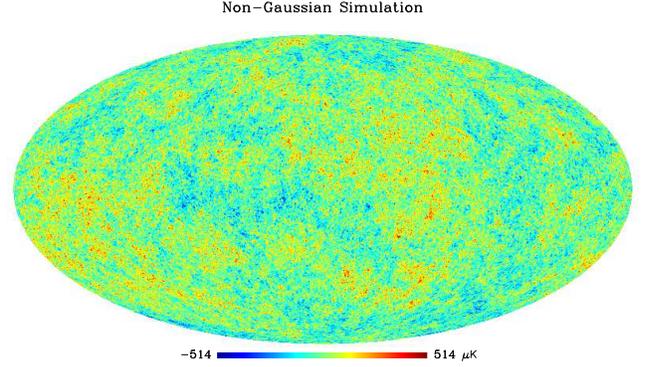}
\end{center}
\caption{A realisation of a non-Gaussian all-sky CMB map 
with a prescribed ensemble-average power spectrum $C_\ell$,
obtained using the non-Gaussian PDF plotted in Fig.~\ref{fig:sho} with
$\alpha_3 = 0.2$, {\sc Healp}ix resolution parameter
$N_{\rmn{side}}= 512$ (WMAP resolution) and $\ell_{\rmn{max}}=1500$.}
\label{fig:ngsimfull}
\end{figure}
\begin{figure}
\begin{center}
\end{center}
\caption{A realisation of a Gaussian all-sky CMB map drawn from the
same ensemble-average power spectrum, $C_\ell$, as the non-Gaussian map
shown in Fig.~\ref{fig:ngsimfull}.}
\label{fig:gsimfull}
\end{figure}
%


The source code to simulate the non-Gaussian CMB maps for both the
full sky and for a small patch of the sky are available at the {\sc NGsims} 
webpage\footnote{http://www.mrao.cam.ac.uk/$\sim$graca/NGsims/}.

As a guide to help the reader reproduce our method, 
we give here a summary of the logical steps to be taken to generate these
non-Gaussian maps (see also documentation at the {\sc NGsims} webpage):

\begin{enumerate}

\item Draw independent identically-distributed pixel values from a
non-Gaussian PDF to create a statistically-isotropic map of non-Gaussian
white noise;

\item Transform the map to harmonic space (with e.g.\ a fast spherical
transform);

\item Scale the harmonic coefficients to enforce the desired power spectrum;

\item Transform back to real space to obtain a non-Gaussian map which has the
desired 2-point correlation function.

\end{enumerate}

We have implemented this method using two classes of non-Gaussian PDF (see documentation in NGsims webpage), 
but for the purposes of illustration, we adopt here PDF (i):

\begin{enumerate}

\item A general PDF based on the energy eigenstates of a linear harmonic
oscillator, which takes the form of a Gaussian multiplied by a series of
Hermite polynomials (see Appendix A).
In principle this expansion can be used to generate any PDF, although if the
expansion is truncated then the available values of the relative skewness are
constrained.  

\item The pixel values are drawn from a Gaussian distribution and then raised
to some (even) integer power.  
This method is very fast to implement and easily generates distributions with
large skewness and so is useful for checking statistical tools.

\end{enumerate}


As we shall see in the next section, the simulation method described
above enables one to generate non-Gaussian maps for which many of the
statistical properties can be calculated analytically.  The range of
possible correlators and polyspectra are, however, rather restricted,
with the scale dependence of the latter controlled solely by the
angular power spectrum (see Section~\ref{sec:poly}).  It is, however,
straightforward to extend our basic method to allow the simulation of
non-Gaussian maps with a much wider range of statistical properties,
as will be discussed in Section~\ref{sec:extsim}.


\section{Statistical properties of the map}
\label{sec:stat_props}

\begin{figure}
\begin{center}
\includegraphics[height = 0.22\textheight]{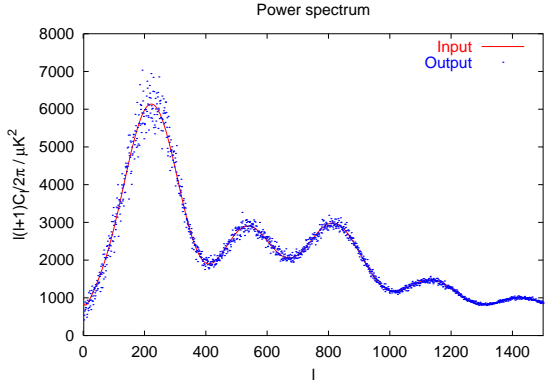}
\end{center}
\caption{The power spectrum of the map shown in Fig.~\ref{fig:ngsimfull}
(points) as compared with the input ensemble-average spectrum
$C_\ell$ (solid line).}
\label{fig:psfull}
\end{figure}

It is clear that, by construction, the non-Gaussian map plotted in
Fig.~\ref{fig:ngsimfull} is statistically isotropic, with an
ensemble-average power spectrum, $C_\ell$, corresponding to a generic
inflationary cosmological model. As a check, the power spectrum of the
map was calculated and found to agree with the input spectrum, as
shown in Fig.~\ref{fig:psfull}.
Owing to the simple manner in which the simulated non-Gaussian map is
produced, many more of its statistical properties may be calculated
analytically, such as the correlators of the pixel values, the
bispectrum and higher-order polyspectra, and the marginalised
probability distribution of the map.  In this section, we again
concentrate on the all-sky case; the corresponding discussion for the
flat-sky approximation is given in Appendix~\ref{sec:flat_sky}.

To facilitate the calculation of the statistical properties of the
map, it is convenient first to express the temperature $t_p$ in each
pixel in terms of the initial pixel values $\{s_p\}$ drawn from the
original non-Gaussian PDF $f_S(s)$. Using equations (\ref{eqn:abar}) and 
(\ref{eqn:tpdef}), we begin by writing
\begin{eqnarray}
t_p & = & \frac{1}{\sqrt{\mu_2 \Omega_{\rm pix}}}
\sum_{\ell, m} \sqrt{C_\ell} a_{\ell m}Y_{\ell m}(\mathbfit{x}_p)
\nonumber \\
& = & \sqrt{\frac{\Omega_{\rm pix}}{\mu_2}}
\sum_{p'} s_{p'} \sum_\ell \sqrt{C_\ell} \sum_m 
Y_{\ell m}(\mathbfit{x}_p)Y^*_{\ell m}(\mathbfit{x}_{p'}),
\label{eqn:tp2}
\end{eqnarray}
where in the second line we have substituted 
for $a_{\ell m}$ from equation (\ref{eqn:almdef}). 
Using the spherical harmonic addition formula 
\begin{equation}
\sum_m  Y_{\ell m}(\mathbfit{x}_p)Y^*_{\ell m}(\mathbfit{x}_{p'})
=\frac{2\ell+1}{4\pi}P_\ell(\mathbfit{x}_p\cdot\mathbfit{x}_{p'}),
\label{eq:addition}
\end{equation}
where $P_\ell(z)$ is a Legendre polynomial, 
we may thus write equation (\ref{eqn:tp2}) as
\be
t_p = \sum_{p'} W_{pp'}s_{p'},
\label{eqn:tpsp}
\ee
where the elements of the weight matrix are given by
\be
W_{pp'} = \sqrt{\frac{\Omega_{\rm pix}}{\mu_2}} \sum_\ell
\frac{2\ell+1}{4\pi} 
\sqrt{C_\ell} P_\ell (\mathbfit{x}_p\cdot\mathbfit{x}_{p'}).
\label{eqn:wdef}
\ee

It is also useful to express the spherical harmonic coefficients,
$\bar{a}_{\ell m}$, of the map as linear superposition of the original 
pixel values $\{s_p\}$. From equations (\ref{eqn:almdef}) and
(\ref{eqn:abar}), we immediately obtain 
\be
\bar{a}_{\ell m} = \sum_p \widetilde{W}_{\ell m, p} s_p,
\label{eqn:abarsp}
\ee
where the elements of this second weight matrix are simply scaled
spherical harmonics evaluated at the $p$th pixel and are given by
\be
\widetilde{W}_{\ell m, p} = \sqrt{\frac{\Omega_{\rm pix}C_\ell}{\mu_2}}
Y^*_{\ell m}(\mathbfit{x}_p).
\label{eqn:w2def}
\ee

Equations (\ref{eqn:tpsp})--(\ref{eqn:w2def}) provide the basic
expressions from which the statistical properties of the 
non-Gaussian map may be obtained.


\subsection{Correlators of the pixel values}

Using equations (\ref{eqn:cuma}) and (\ref{eqn:tpsp}), a 
general expression for the connected $n$-point correlator of the pixel
values in the non-Gaussian map is given by
\begin{eqnarray}
\langle t_{p_1} \cdots t_{p_n} \rangle_c
& = &  \sum_{q_1 \cdots q_n} W_{p_1q_1}\cdots W_{p_nq_n}
\langle  s_{q_1} \cdots s_{q_n} \rangle_c \nonumber \\
& = & \kappa_n \sum_{q} W_{p_1 q} \cdots W_{p_nq}, 
\label{eq:correlators}
\end{eqnarray}
where $\kappa_n$ is the $n$th cumulant of $f_S(s)$.
Inserting the expression (\ref{eqn:wdef}) into this result, we obtain
\begin{equation}
\langle t_{p_1} \cdots t_{p_n} \rangle_c 
= 
\frac{\kappa_n}{\Omega_{\rm pix}} \!
\left(\frac{\Omega_{\rm pix}}{\mu_2}\right)^{n/2}\!\!\!
\sum_{\ell_1\cdots\ell_n} \!\!\alpha_{\ell_1\cdots\ell_n}
I_{\ell_1\cdots\ell_n}(\mathbfit{x}_{p_1}, \ldots, \mathbfit{x}_{p_n}),
\end{equation}
where we have defined the quantities
\begin{eqnarray}
\alpha_{\ell_1\cdots\ell_n} \!\!\!\! & \equiv & \!\!\!\!
\frac{(2\ell_1+1)\cdots (2\ell_n+1)}{(4\pi)^n}
\sqrt{C_{\ell_1}\cdots C_{\ell_n}}, \\
I_{\ell_1\cdots \ell_n}(\mathbfit{x}_{p_1},\ldots, \mathbfit{x}_{p_n})
\!\!\!\! & \equiv & \!\!\!\!
\sum_q \! P_{\ell_1}(\mathbfit{x}_{p_1}\cdot\mathbfit{x}_q)\cdots
P_{\ell_n}(\mathbfit{x}_{p_n}\cdot\mathbfit{x}_q)\,\Omega_{\rm pix}.
\end{eqnarray}
Taking the continuum limit of the sum over $q$, we see that an analytic
expression for the $n$-point correlator may be obtained by evaluating
\begin{equation}
I_{\ell_1\cdots \ell_n}(\mathbfit{x}_{p_1},\ldots, \mathbfit{x}_{p_n})  \approx 
\int_{4\pi}\!\! 
 P_{\ell_1}(\mathbfit{x}_{p_1}\cdot\mathbfit{x})\cdots
P_{\ell_n}(\mathbfit{x}_{p_n}\cdot\mathbfit{x})\,\rmn{d}\Omega.
\end{equation}
Clearly, $I_{\ell_1\cdots \ell_n}$ is invariant under rigid rotations of its
vector arguments, and this property is inherited by the $n$-point
correlator as required by statistical isotropy. The integral
may be related to the $n$-polar harmonics with zero total angular
momentum~\citep{Varshalovich:1988}; this reduction for the first
few values of $n$ is described further in Appendix~\ref{sec:integrals}. 

Using the results derived in Appendix~\ref{sec:integrals}, one
recovers, for example, the well-known result for the 2-point
correlator
\begin{equation}
\langle t_{p_1} t_{p_2} \rangle  = \sum_\ell \frac{2\ell+1}{4\pi}
C_\ell P_\ell(\mathbfit{x}_{p_1}\cdot\mathbfit{x}_{p_2}),
\label{eq:2point}
\end{equation}
and an explicit form for the 3-point correlator given by
\[
\langle t_{p_1} t_{p_2} t_{p_3} \rangle_c 
\]
\vspace*{-2mm}
\begin{eqnarray}
&=& \!\!\!\frac{\mu_3}{\mu_{2}^{3/2}} \Omega_{\rmn{pix}}^{1/2} 
\sum_{\ell_1,\ell_2,\ell_3} 
\sqrt{\frac{D_{\ell_1}D_{\ell_2} D_{\ell_3}}{4\pi}} 
\left(\begin{array}{ccc}\ell_{1} & \ell_{2} & \ell_{3} \\
	            0 & 0 & 0
\end{array}
\right) \nonumber \\ 
&\times& 
\!\!\!\sum_{m_1 m_2 m_3}
\left( 
\begin{array}{ccc}\ell_{1} & \ell_{2} & \ell_{3}\\
	          m_{1} & m_{2} & m_{3}
\end{array}
\right)
Y^*_{\ell_1 m_1}(\mathbfit{x}_{p_1}) 
Y^*_{\ell_2 m_2}(\mathbfit{x}_{p_2}) 
Y^*_{\ell_3 m_3}(\mathbfit{x}_{p_3}) \nonumber,
\end{eqnarray}
in which we have defined $D_\ell \equiv (2\ell+1) C_\ell$.


\subsection{Bispectrum and higher-order polyspectra}
\label{sec:poly}

To compute the polyspectra of the non-Gaussian map we form the
$n$th connected correlator of its multipoles using 
equations~(\ref{eqn:cuma}) and (\ref{eqn:abarsp}):
\begin{eqnarray}
\langle \bar{a}_{\ell_1 m_1}\cdots\bar{a}_{\ell_n m_n} \rangle_c
& = &  \sum_{p_1 \cdots p_n} \widetilde{W}_{\ell_1 m_1,p_1}\cdots 
\widetilde{W}_{\ell_n m_n,p_n}
\langle  s_{p_1} \cdots s_{p_n} \rangle_c \nonumber \\
& = & \kappa_n \sum_{p}\widetilde{W}_{\ell_1 m_1,p}\cdots 
\widetilde{W}_{\ell_n m_n,p}.
\end{eqnarray}
Inserting the expression (\ref{eqn:w2def}) into this
result, we obtain
\be
\langle \bar{a}_{\ell_1 m_1}\cdots\bar{a}_{\ell_n m_n} \rangle_c
= \frac{\kappa_n}{\Omega_{\rm pix}} \!
\left(\frac{\Omega_{\rm pix}}{\mu_2}\right)^{n/2}
\sqrt{C_{\ell_1}\cdots C_{\ell_n}} \, J_{\ell_1 m_1,\ldots,\ell_n m_n},
\label{eqn:polyspec}
\ee
where 
\begin{eqnarray}
J_{\ell_1 m_1,\ldots,\ell_n m_n} & = & 
\sum_p Y^*_{\ell_1 m_1}(\mathbfit{x}_p)\cdots
Y^*_{\ell_n m_n}(\mathbfit{x}_p)\,\Omega_{\rm pix} \nonumber \\
& \approx &
\int_{4\pi} Y^*_{\ell_1 m_1}(\mathbfit{x})\cdots
Y^*_{\ell_n m_n}(\mathbfit{x})\,\rmn{d}\Omega.
\end{eqnarray}
Integrals of this form may be reduced to products of 3$j$ symbols
as discussed in Appendix~\ref{sec:integrals}.

For the case $n=2$, one immediately recovers
$\langle \bar{a}_{\ell m} \bar{a}^*_{\ell' m'} \rangle = C_\ell
\delta_{\ell\ell'}\delta_{mm'}$. For $n=3$, we find that
\begin{equation}
\langle \bar{a}_{\ell_1 m_1}\bar{a}_{\ell_2 m_2}\bar{a}_{\ell_3 m_3} \rangle
= \left(\begin{array}{ccc}
\ell_1 & \ell_2 & \ell_3 \\
m_1 & m_2 & m_3 
\end{array}
\right) B_{\ell_1 \ell_2 \ell_3},
\end{equation}
where the bispectrum coefficients of the simulated non-Gaussian map
are given by
\begin{equation}
B_{\ell_1 \ell_2 \ell_3} 
= \frac{\kappa_3}{\Omega_{\rm pix}} \!
\left(\frac{\Omega_{\rm pix}}{\mu_2}\right)^{3/2}
\sqrt{\frac{D_{\ell_1}D_{\ell_2} D_{\ell_3}}{4\pi}} 
\left(\begin{array}{ccc}
\ell_1 & \ell_2 & \ell_3 \\
0 & 0 & 0 
\end{array}
\right).
\end{equation}
It is convenient also to introduce the `normalised' (by the power
spectrum) reduced bispectrum $\hat{b}_{\ell_1 \ell_2 \ell_3}$ by the
relation
\begin{equation}
B_{\ell_1 \ell_2 \ell_3} = 
\sqrt{\frac{D_{\ell_1}D_{\ell_2} D_{\ell_3}}{4\pi}} 
\left(\begin{array}{ccc}
\ell_1 & \ell_2 & \ell_3 \\
0 & 0 & 0 
\end{array}
\right)\hat{b}_{\ell_1 \ell_2 \ell_3}.
\label{bisresult}
\end{equation}
Thus, for our simulated map, $\hat{b}_{\ell_1 \ell_2 \ell_3}$ has a
constant value given by
\begin{equation}
\hat{b}_{\ell_1 \ell_2 \ell_3} = \frac{\kappa_3}{\Omega_{\rm pix}} \!
\left(\frac{\Omega_{\rm pix}}{\mu_2}\right)^{3/2}.
\label{eq:bispecpred}
\end{equation}

We see that the amplitude of the bispectrum is determined by the
variance $\mu_2$ and skewness $\kappa_3$ of the original non-Gaussian PDF
$f_S(s)$. If $f_S(s)$ were Gaussian, for example, the bispectrum would
clearly vanish. Even if $f_S(s)$ has a non-zero skewness, however, we
note from equation~(\ref{eqn:polyspec}) that, in the limit that the number of
pixels tends to infinity, all polyspectra with $n \geq 3$ vanish.
This is a consequence of each pixel being the weighted sum of independent
identically-distributed variates, so that in the limit of an infinite
number of pixels the processed map $T(\mathbfit{x})$ tends to Gaussian
by the central-limit theorem. Fortunately, in practice,
we can bypass this property as $N_{\rmn{pix}}$ increases by scaling the
cumulants of the initial distribution $f_S(s)$. For example, to get a
bispectrum independent of $N_{\rmn{pix}}$, we must scale $\kappa_3$ such that
\begin{equation}
\frac{\kappa_3}{\Omega_{\rmn{pix}}} \!
\left(\frac{\Omega_{\rmn{pix}}}{\mu_2}\right)^{3/2} = \mbox{constant}.
\end{equation}
As mentioned in Appendix~\ref{sec:sho}, for the particular non-Gaussian PDF
used here, generating higher values of the relative skewness requires
one to have a larger range of the generalized cumulants parameters $\alpha_n$
non zero.

\begin{figure}
\begin{center}
\includegraphics[height = 0.22\textheight]{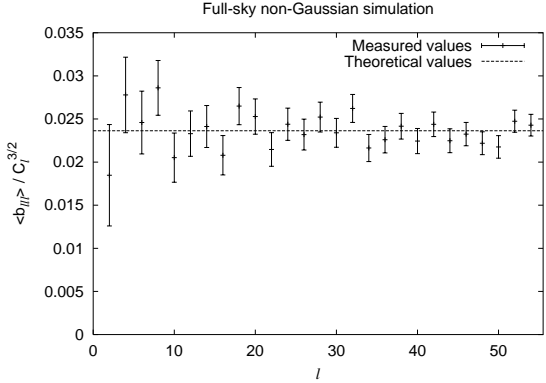}
\end{center}
\caption{Non-zero diagonal components of the bispectrum estimated from 
$300\,000$
non-Gaussian simulations with the same parameters as those used in
Fig.~\ref{fig:ngsimfull}. The mean over the simulations and its standard error
are plotted. The dashed line is the theoretical
ensemble-average value. Note that the diagonal bispectrum
necessarily vanishes for odd $\ell$  by parity.}
\label{fig:fullbiskew}
\end{figure}

As a check on our calculations, the normalised reduced bispectra of an
ensemble of $300\,000$ non-Gaussian maps was calculated using an 
estimator defined in \cite{spergel} as
\begin{eqnarray}
\hat{b}_{\ell_1 \ell_2 \ell_3} &=& \frac{1}{4 \pi}\int e_{\ell_1}(\mathbfit{x}) e_{\ell_2}(\mathbfit{x}) e_{\ell_3}(\mathbfit{x}) \rmn{d}\Omega \nonumber \\
& & \qquad\times  \sqrt{\frac{4 \pi}{D_{\ell_1} D_{\ell_2} D_{\ell_3}}} 
\left( 
\begin{array}{ccc}\ell_1 & \ell_2 & \ell_3\\
	           0 & 0 &  0
\end{array}
\right)^{-2},
\label{bisest}
\end{eqnarray} 
where, $e_{\ell} (\mathbfit{x}) =\sqrt{\frac{4 \pi}{2 \ell +1}} \sum_m a_{\ell m} Y_{\ell m} (\mathbfit{x})$. The
resulting mean values of $\hat{b}_{\ell \ell \ell}$ are plotted in
Fig.~\ref{fig:fullbiskew}, together with the associated uncertainties.
Note that it is important to divide the measured bispectrum by the ensemble-averaged values of the power spectrum; dividing by the measured values for each simulation produces a biased estimator.
The predicted value of $\hat{b}_{\ell_1 \ell_2 \ell_3}$ was calculated
using (\ref{eq:bispecpred}) and is plotted as the solid line in the
figure.  We see that the measured and predicted values are fully
consistent.

Returning to equation~(\ref{eqn:polyspec}), if one considers $n=4$ and
follows the notation of \cite{hu}, one finds that the connected part
of the trispectrum of the non-Gaussian simulation is given by
\begin{eqnarray}
T_{\ell_3 \ell_4}^{\ell_1 \ell_2} (L)
& = & \frac{\Omega_{\rmn{pix}} \kappa_4}{\mu_{2}^{2}} 
\sqrt{D_{\ell_1} D_{\ell_2}D_{\ell_3}D_{\ell_4}} 
\frac{2L +1}{4 \pi} \nonumber \\
& & \qquad\times\left(
\begin{array}{ccc} \ell_{1} & \ell_{2} & L\\
		   0 & 0 & 0 
\end{array}
\right)
\left(
\begin{array}{ccc} \ell_{3} & \ell_{4} & L\\
		   0 & 0 & 0 
\end{array}
\right).   
\end{eqnarray}
Analytic expressions for higher-order polyspectra may be obtained in
an analogous manner.


\subsection{Cumulants and marginal distribution}
\label{sec:cumulants}

Finally, we consider the marginal PDF $f_T(t)$ of the processed map.
This is mostly easily carried out by considering the cumulants of
$f_T(t)$, and relating them to those of the original non-Gaussian PDF
$f_S(s)$. 

The cumulants of the marginal PDF are equal to the connected parts of the
correlators of the pixel temperatures with all pixels in the correlator
the same. From equation~(\ref{eq:correlators}), we find that
\begin{eqnarray}
\kappa'_n &=& \left(\sum_{p'}^{N_{\rmn{pix}}} W^n_{pp'}\right) \kappa_n
\nonumber \\
	  &=& \frac{\kappa_n}{\Omega_{\rmn{pix}}} \left(
\frac{\Omega_{\rmn{pix}}}{\mu_2}\right)^{n/2}\sum_{\ell_1 \cdots \ell_n}
\alpha_{\ell_1\cdots \ell_n}I_{\ell_1\cdots \ell_n}
(\mathbfit{x}_p,\ldots,\mathbfit{x}_p),
\label{eqn:cum}
\end{eqnarray}
which is independent of the choice of pixel $p$. Evaluating
the rotationally-invariant function
$I_{\ell_1\cdots \ell_n}(\mathbfit{x}_p,\ldots, \mathbfit{x}_p)$ along the polar
axis, where $Y_{\ell m}(\hat{\mathbfit{z}})=\sqrt{(2l+1)/4\pi}\delta_{m0}$, we
find that the first few cumulants (beyond
$\kappa'_1=0$) are
\begin{eqnarray*}
\kappa'_{2}\!\!\!\! & = & \!\!\!\!
\sum_{\ell} \frac{2 \ell +1}{4 \pi} C_{\ell} \\
\kappa'_{3} \!\!\!\! & = & \!\!\!\! \sum_{\ell_{1} \ell_{2} \ell_{3}} 
B_{\ell_{1} \ell_{2} \ell_{3}} 
\sqrt{\frac{2 \ell_{1} +1}{4\pi}\cdots\frac{2 \ell_{3} +1}{4 \pi}}
\left(
\begin{array}{ccc} \ell_{1} & \ell_{2} & \ell_{3}\\
		   0 & 0 & 0 
\end{array}
\right) \\  
\kappa'_{4} \!\!\!\! & = & \!\!\!\! \sum_{\ell_{1} \cdots \ell_{4}} 
\sqrt{\frac{2 \ell_{1} +1}{4 \pi} \cdots \frac{2 \ell_{4} +1}{4 \pi}} \\
& & 
\qquad \times \sum_{L} T_{\ell_3 \ell_4}^{\ell_1 \ell_2} (L)
\left( 
\begin{array}{ccc} \ell_{1} & \ell_{2} & L\\
		   0 & 0 & 0 
\end{array}
\right)
\left(
\begin{array}{ccc} \ell_{3} & \ell_{4} & L\\
		   0 & 0 & 0 
\end{array}
\right),   
\end{eqnarray*}
where we have written the results in such as way that they are true
generally, for any statistically-isotropic map.

\begin{figure}
\begin{center}
\includegraphics[height= 0.3\textheight,angle = 270]
{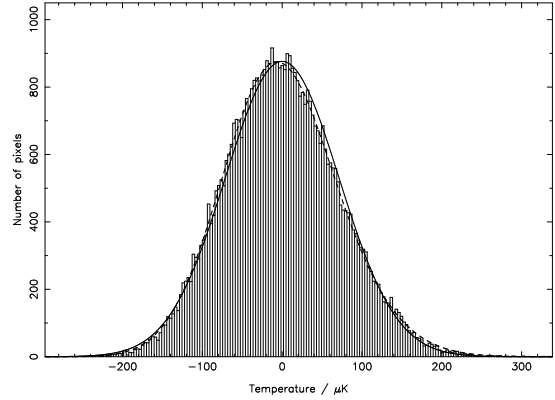}
\end{center}
\caption{Histogram of pixel temperatures for the final processed non-Gaussian map for 
$N_{side}=64$ and $\ell_{\rmn{max}}=128$, created using an initial PDF with 
parameters $\alpha_{3}=0.27$ and $\sigma_{0}=1$.
Also shown are the Edgeworth expansion for $f_T(t)$ (dashed
line), given by equation~(\ref{eqn:edgeworth}) but truncated at $n=4$, and a
Gaussian with the same variance (solid line).}
\label{fig:ngsimhist}
\end{figure}

In principle, knowledge of the complete set of 
cumulants $\kappa'_n$ may be used to obtain an
explicit expression for the marginal PDF $f_T(t)$. This could be
carried out, for example, by first obtaining its moment-generating function 
\begin{equation}
M_T(k) = \langle \rmn{e}^{\rmn{i}k t} \rangle =
\exp\left(\sum_{n=1}^\infty \frac{(\rmn{i}k)^n}{n!}\kappa'_n\right), 
\end{equation}
and then performing an inverse Fourier transform to yield $f_T(t)$.
Alternatively, for a weakly non-Gaussian distribution,
one can employ the Edgeworth expansion.
In this approach, the PDF is expressed as an asymptotic expansion
around a Gaussian with mean zero and variance $\sigma^2=\kappa'_2$ to yield
\begin{eqnarray}
f_T(t) & \approx &
\frac{\rmn{e}^{-\hat{t}^2}}{\sqrt{2 \pi\sigma^{2}}} \Bigl(1
+ \frac{\kappa'_3/\sigma^3}{12 \sqrt{2}}H_3(\hat{t})+
\frac{\kappa'_4/\sigma^4}{96} H_4(\hat{t}) \nonumber \\
&&
+ \frac{\kappa'_5/\sigma^5}{480 \sqrt{2}} H_5 (\hat{t})
+ \frac{(\kappa'_6 + 10 {\kappa'_3}^{2})/\sigma^6}{5760} H_6
(\hat{t}) + \cdots  \Bigr), 
\label{eqn:edgeworth}
\end{eqnarray}
where $\hat{t}=t/\sqrt{2}\sigma$. 
When $\kappa'_n/\sigma^n$ is small for $n$ larger than some integer
one can use a finite number of cumulants as an
acceptable approximation to the distribution. As pointed out by
\citet{rocha}, however, this might no longer be a PDF and, in
particular, might deviate from the original distribution in the
tails. In Fig.\ref{fig:ngsimhist} we plot the histogram of the pixel temperatures for a
non-Gaussian map with $N_{\rmn{side}}=64$ and $\ell_{\rmn{max}}=128$
(somewhat lower resolution than that plotted in Fig.~\ref{fig:ngsimfull}), generated
from an initial PDF with parameters $\alpha_3=0.27$
and $\sigma_0=1$. Overplotted is the Edgeworth expansion of $f_T(t)$,
equation~(\ref{eqn:edgeworth}), truncated at $n=4$. The cumulants of $f_T(t)$ can be
computed efficiently from the first expression in equation~(\ref{eqn:cum}); we
find $\kappa'_1 = 0$, $\kappa'_2 \approx 13809\, {\mu\rmn{K}}^{2}$, $\kappa'_3
\approx 177234\, {\mu\rmn{K}}^{3}$ and $\kappa'_4 \approx 6307963 \, {\mu\rmn{K}}^{4}$.
The Edgeworth expansion agrees with the simulation results better than a
Gaussian with the same variance.
%


\section{Extended simulation method}
\label{sec:extsim}

We see from the previous section that our basic simulation method
generates maps with a rather restricted range of possible correlators
and polyspectra, with the scale dependence of the latter controlled
solely by the angular power spectrum.  It is straightforward, however,
to extend our basic method to allow the simulation of non-Gaussian
maps with a much wider range of statistical properties.  In
particular, there is no fundamental requirement for the method to be
restricted to the same set of cumulants over the whole range of
scales.

The procedure is as follows. One divides the range of multipoles,
$\ell$, into non-overlapping bins. A given bin $B$ will be a set
$B=[\ell_{B_{\rmn{min}}},\ell_{B_{\rmn{max}}}]$.  For each bin $B$ we
simulate a map $t_{Bp}$, with non-zero $C_\ell$ only for $\ell \in B$
and a non-Gaussian PDF that can differ between bins.  The final map is
a superposition of these band maps, i.e. with pixel values 
\be t_p =
\sum_{B} t_{Bp}.
\label{summap}
\ee 
Since each map $t_{Bp}$ is individually statistically isotropic,
then so too is their sum.  Moreover, from equation~(\ref{bisresult}), we see
that the bispectrum for each band map can only be
non-zero within the corresponding bin (this is also true for
the connected parts of higher-order polyspectra).  
The $n$th cumulant of the final map is also simply the sum of the $n$th cumulants
of the individual band maps. With this
method we are thus able to generate maps with more general statistical
properties.  

\begin{figure}
\begin{center}
\includegraphics[height= 0.35\textheight,angle = 90]
{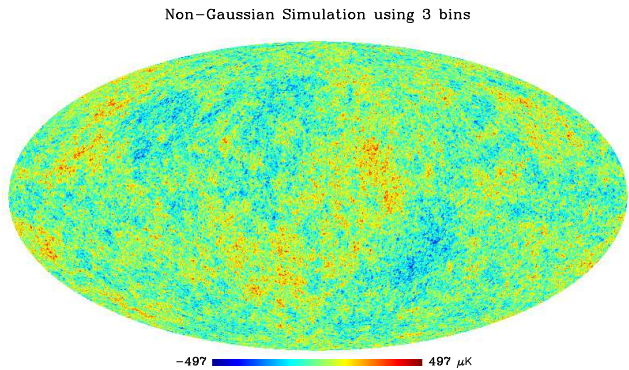}
\end{center}
\caption{A realisation of a non-Gaussian all-sky map 
simulation using three
bins: $B_1=[0,500],\alpha_3=0.1$;
$B_2=[501,1000],\alpha_3=0.2$; and 
$B_3=[1001,1500],\alpha_3=0.25$.
}
\label{fig:ngsim3bin}
\end{figure}

\begin{figure}
\begin{center}
\includegraphics[height = 0.22\textheight]{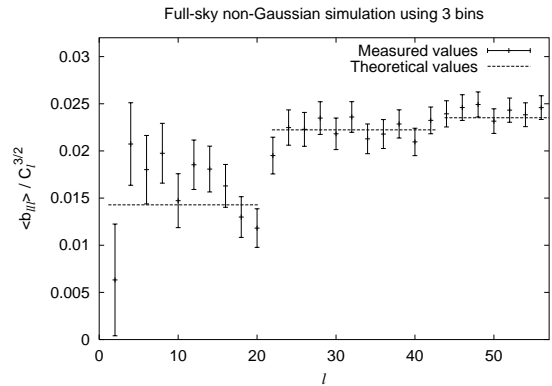}
\end{center}
\caption{Non-zero diagonal components of the bispectrum estimated from 
$300\,000$
non-Gaussian simulations using three
bins: $B_1=[0,21],\alpha_3=0.1$;
$B_2=[22,43],\alpha_3=0.2$; and 
$B_3=[44,56],\alpha_3=0.25$. The mean over the simulations and its standard error
are plotted. The dashed lines shows the theoretical ensemble-average value in each bin.}
\label{fig:full3binbiskew}
\end{figure}

For example, by choosing appropriate values of $\kappa_3$ for the
non-Gaussian PDF used to simulate each band map, one can arrange for
the summed map (\ref{summap}) to have a given power spectrum $C_\ell$
and an arbitrary prescribed constant value of the reduced normalised
bispectrum $\hat{b}_{\ell_1\ell_2\ell_3}$ in each bin. As an
illustration, in Fig.~\ref{fig:ngsim3bin} we plot a non-Gaussian map
generated using three bins. 
In each bin, the non-Gaussian PDF used was of the form
given in equation~(\ref{like}) with $\sigma_{0}=1$. 
However, the values of $\alpha_3$ used in each bin
were $\alpha_{3}=0.1$, $0.2$ and $0.25$ respectively.
As a check on our calculations, the normalised reduced bispectrum of an
ensemble of $300\,000$ such non-Gaussian maps was calculated using the
estimator (\ref{bisest}). The resulting mean values of $\hat{b}_{\ell \ell
\ell}$, for individual values of $\ell$, are
plotted in Fig.~\ref{fig:full3binbiskew}, together with the associated
uncertainties.  The predicted value of $\hat{b}_{\ell \ell \ell}$
in each of the three broad bins was calculated using equation~(\ref{eq:bispecpred})
 and are plotted as the dashed lines in the
figure.  We see that once again the measured and predicted values are
fully consistent.

How general is our extended method? We can in principle efficiently
generate maps with arbitrary diagonal bispectra, i.e.\ with any given
$C_\ell$ and $B_{\ell\ell\ell}$.  This is of some importance because
currently known primordial theories of non-Gaussianity lead to more
general combinations of angular power spectra and bispectra than can
be created from our method with a single univariate PDF $f_S(s)$. It
is not possible, however, to generate specific models of primordial
non-Gaussianity exactly with our extended method.  To do that one
would have to be able to choose arbitrary values for the angular
spectrum and for all components of the bispectrum (and higher
polyspectra). This would involve going beyond the simple one-point PDF
methods advocated here.


\section{Conclusions}
\label{sec:conc}

We presented a simple, fast method for simulating
statistically-isotropic non-Gaussian CMB maps with a given power
spectrum and analytically calculable bispectrum.  We showed that our
technique allows one to describe the statistical properties of the map
by computing analytically the $n$th-order polyspectra, and the
$n$th-order correlators of the pixel values. We showed that these can be
expressed in terms of the $n$-polar harmonics with zero total angular
momentum, and we describe this reduction for the first few values of
$n$.  We also recovered analytically the one-dimensional marginalised
distribution function in terms of its cumulants.  The univariate
non-Gaussian distribution, from which the pixel values are drawn
independently in the first stage of the simulation process, fully
determines the statistical properties of the final map.  Here we used a
non-Gaussian distribution derived from the wavefunctions of the
harmonic oscillator.  Simulations of both the full sky and a small
patch of the sky were generated and corresponding statistical analysis
performed.  As a check on our calculations we computed both the power
spectrum and bispectrum of the simulated maps and found them to be fully consistent.
  The simulation method described here clearly
enables one to generate maps with well-defined correlators and
polyspectra. We extended the method to encompass different set of
cumulants over the whole range of scales, generating maps with
arbitrary power spectra and diagonal bispectra for different scales.  
It is not possible, however, to generate specific models of primordial
non-Gaussianity exactly with our extended method, since these require
off-diagonal bispectrum coefficients to be specified arbitrarily.
This would involve going beyond the simple one-point PDF
methods advocated here.

The source code to simulate the non-Gaussian CMB maps for both the
full sky and for a small patch of the sky are available at the {\sc NGsims} 
webpage\footnote{http://www.mrao.cam.ac.uk/$\sim$graca/NGsims/}.

A pertinent question is what other statistical properties can be
calculated analytically for the class of non-Gaussian maps we have
investigated. Of particular interest are the phase associations
between different harmonic coefficients. In \cite{matsubara}, a
general relationship between phase correlations and the hierarchy of
polyspectra in Fourier space is established. It is also stated
that the phase correlations are related to the polyspectra through the
non-uniform distribution of the phase sum $\theta_{\mathbfit{k}_1}+ \theta_{\mathbfit{k}_2}+
\cdots + \theta_{\mathbfit{k}_N}$ with closed vectors $\mathbfit{k}_1+ \mathbfit{k}_2+
\cdots + \mathbfit{k}_N=\mathbfit{0}$. We are currently investigating the form of
the distribution function of this phase sum in our maps. A study of
the Minkowski functionals of our non-Gaussian maps is also underway.
 

\section{Ackowledgements}
We thank Carlo Contaldi and Neil Turok for useful discussions, and
Martin Kunz and Grazia De Troia for providing us with the bispectrum code for the full-sky
case.  
Some of the results in this paper have
been derived using the {\sc Healp}ix (Gorski, Hivon, and Wandelt 1999) package.
GR acknowledges a Leverhulme Fellowship at the University of
Cambridge. PF and AC acknowledge Royal Society University Research
Fellowships. SS acknowledges support by a PPARC studentship.

\bibliography{ng_sims}

\begin{thebibliography}{}

\bibitem[\protect\citeauthoryear{Acquaviva, Bartolo, Matarrese \&
  Riotto}{Acquaviva et~al.}{2003}]{Acquaviva}
Acquaviva V.,  Bartolo N.,  Matarrese S.,    Riotto A.,  2003, Nuclear Physics
  B, 667, 119

\bibitem[\protect\citeauthoryear{Bartolo, Komatsu, Matarrese \& Riotto}{Bartolo
  et~al.}{2004}]{bartolo04}
Bartolo N.,  Komatsu E.,  Matarrese S.,    Riotto A.,  2004, Physics Reports

\bibitem[\protect\citeauthoryear{Bartolo, Matarrese \& Riotto}{Bartolo
  et~al.}{2004}]{Bartolo}
Bartolo N.,  Matarrese S.,    Riotto A.,  2004, J.\ High Energy Phys., 04, 006

\bibitem[\protect\citeauthoryear{Contaldi, Bean \& Magueijo}{Contaldi
  et~al.}{1999}]{joao}
Contaldi C.,  Bean R.,    Magueijo J.,  1999, Phys. Lett., B468, 189

\bibitem[\protect\citeauthoryear{Contaldi \& Magueijo}{Contaldi \&
  Magueijo}{2001}]{contaldi}
Contaldi C.,  Magueijo J.,  2001, Phys. Rev., D63, 103512

\bibitem[\protect\citeauthoryear{Crittenden}{Crittenden}{2000}]{igloo}
Crittenden R.~G.,  2000, Astrophysical Letters and Communications, 37, 377

\bibitem[\protect\citeauthoryear{Dvali, Gruzinov \& Zaldarriaga}{Dvali
  et~al.}{2004}]{dvali04}
Dvali G.,  Gruzinov A.,    Zaldarriaga M.,  2004, Phys.\ Rev.\, D69, 023505

\bibitem[\protect\citeauthoryear{Edmonds}{Edmonds}{1974}]{Edmonds:1974}
Edmonds A.~R.,  1974, Angular Momentum in Quantum Mechanics.
Princeton University Press, Princeton, New Jersey

\bibitem[\protect\citeauthoryear{Gangui, Martin \& Sakellarioudou}{Gangui
  et~al.}{2002}]{gangui}
Gangui A.,  Martin J.,    Sakellarioudou M.,  2002, Phys. Rev. D, 66, 083502

\bibitem[\protect\citeauthoryear{Gangui, Pogosian \& Winitzki}{Gangui
  et~al.}{2002}]{gangui01}
Gangui A.,  Pogosian L.,    Winitzki S.,  2002, New Astronomy Reviews, 46, 681

\bibitem[\protect\citeauthoryear{G\'{o}rski, Hivon \& Wandelt}{G\'{o}rski
  et~al.}{1999}]{healpix}
G\'{o}rski K.~M.,  Hivon E.,    Wandelt B.~D.,  1999, in Banday A.~J.,  Sheth
  R.~S.,   Costa L.~D.,  eds, Proceedings of the MPA/ESO Cosmology Conference
  `Evolution of Large-Scale Structure' PrintPartners Ipskamp, NL, pp 37--42

\bibitem[\protect\citeauthoryear{Hu}{Hu}{2001}]{hu}
Hu W.,  2001, Phys. Rev., D64, 083005

\bibitem[\protect\citeauthoryear{Komatsu et~al.,}{Komatsu
  et~al.}{2003}]{komatsu}
Komatsu E.,  et~al., 2003, ApJS, 148, 135H

\bibitem[\protect\citeauthoryear{Liguori, Matarrese \& Moscardini}{Liguori
  et~al.}{2003}]{liguori}
Liguori M.,  Matarrese S.,    Moscardini L.,  2003, ApJ., 597, 57

\bibitem[\protect\citeauthoryear{Lyth \& Wands}{Lyth \& Wands}{2002}]{lyth02}
Lyth D.~H.,  Wands D.,  2002, Phys.\ Lett.\ B, 524, 5

\bibitem[\protect\citeauthoryear{Ma}{Ma}{1985}]{Ma:1985}
Ma S.~K.,  1985, Statistical Mechanics.
World Scientific, Philadelphia

\bibitem[\protect\citeauthoryear{Maldacena}{Maldacena}{2003}]{Maldacena}
Maldacena J.,  2003, J.\ High Energy Phys., 05, 013

\bibitem[\protect\citeauthoryear{Martin, Riazuelo \& Sakellarioudou}{Martin
  et~al.}{2000}]{martin}
Martin J.,  Riazuelo A.,    Sakellarioudou M.,  2000, Phys. Rev. D, 61, 083518

\bibitem[\protect\citeauthoryear{Mart\'{\i}nez-Gonz\'{a}lez, Gallegos, Argueso,
  Cay\'{o}n \& Sanz}{Mart\'{\i}nez-Gonz\'{a}lez et~al.}{2002}]{Martinez}
Mart\'{\i}nez-Gonz\'{a}lez E.,  Gallegos J.,  Argueso F.,  Cay\'{o}n L.,
  Sanz J.,  2002, MNRAS, 336, 22

\bibitem[\protect\citeauthoryear{Matsubara}{Matsubara}{2003}]{matsubara}
Matsubara T.,  2003, ApJ., 591, L79

\bibitem[\protect\citeauthoryear{Rocha, Magueijo, Hobson \& Lasenby}{Rocha
  et~al.}{2001}]{rocha}
Rocha G.,  Magueijo J.,  Hobson M.,    Lasenby A.,  2001, Phys. Rev., D64,
  063512

\bibitem[\protect\citeauthoryear{Smith et~al.,}{Smith  et~al.}{2004}]{smith04}
Smith S.,  et~al., 2004, MNRAS, 352, 887

\bibitem[\protect\citeauthoryear{Spergel \& Goldberg}{Spergel \&
  Goldberg}{1999}]{spergel}
Spergel D.~N.,  Goldberg D.~M.,  1999, Phys. Rev., D59, 103001

\bibitem[\protect\citeauthoryear{Varshalovich, Moskalev \&
  Khersonskii}{Varshalovich et~al.}{1988}]{Varshalovich:1988}
Varshalovich D.~A.,  Moskalev A.~N.,    Khersonskii V.~K.,  1988, Quantum
  Theory of Angular Momentum.
World Scientific, Singapore

\bibitem[\protect\citeauthoryear{Vio, Andeani, Tenorio \& Wamsteker}{Vio
  et~al.}{2001}]{vio01}
Vio R.,  Andeani P.,  Tenorio L.,    Wamsteker W.,  2001, PASP, 113, 1009

\bibitem[\protect\citeauthoryear{Vio, Andeani, Tenorio \& Wamsteker}{Vio
  et~al.}{2002}]{vio02}
Vio R.,  Andeani P.,  Tenorio L.,    Wamsteker W.,  2002, PASP, 114, 1281

\bibitem[\protect\citeauthoryear{Zaldarriaga}{Zaldarriaga}{2000}]{zaldarriaga0%
0}
Zaldarriaga M.,  2000, Phys.\ Rev.\, D62, 063510

\end{thebibliography}
\bibliographystyle{mn2e}
\bsp

\appendix
\onecolumn

\section{Non-Gaussian PDFs based on the harmonic oscillator}
\label{sec:sho}

In this appendix, we summarise the class of probability distribution
functions (PDFs) derived from the Hilbert space of a linear harmonic
oscillator, which was developed by \cite{rocha}. The original
non-Gaussian distribution, $f_S(s)$, used in the main text to produce
the simulated non-Gaussian maps is an example of such a PDF.

This general PDF is based on the coordinate-space wavefunctions of the
energy eigenstates of a linear harmonic oscillator, and takes the form
of a Gaussian multiplied by the square of a (possibly finite) series
of Hermite polynomials whose coefficients $\alpha_{n}$ are used as
non-Gaussian qualifiers. In particular, if $x$ is a general random
variable, the most general PDF has the form
\begin{equation} 
p(x)=|\psi|^2= \rmn{e}^{-x^2/(2 \sigma_0^2)}
\left|\sum_n
\alpha_n C_n H_n{\left(x\over {\sqrt 2}\sigma_0\right)}\right|^2,
\end{equation}
where $H_n(x)$ are the Hermite polynomials, and the quantity
$\sigma_0^2$ is the variance associated with the (Gaussian)
probability distribution for the ground state $|\psi_0|^2$. The
constants $C_n$ are fixed by normalising the individual states.  The
only constraint upon the amplitudes $\alpha_n$ is
\begin{equation}\label{const}     
\sum |\alpha_n|^2 = 1.
\end{equation}
This is a simple algebraic expression which can be eliminated
explicitly by writing $\alpha_0= \sqrt{1- \sum_1^\infty |\alpha_n|^2
}$.  Thus the coefficients $\alpha_{n}$ can be independently set to
zero without mathematical inconsistency \citep{rocha}. Moreover, these
coefficients can be written as series of cumulants \citep*{joao} and
should indeed be regarded as non-perturbative generalisations of
cumulants.

For the simulations in the main text, we use the
non-Gaussian PDF for which all $\alpha_n$ are set to
zero, except for the real part of $\alpha_3$ (and consequently
$\alpha_{0}$).  The reason for this choice 
is that this quantity reduces to
the skewness in the perturbative regime.  The imaginary part of
$\alpha_3$ is only meaningful in the non-perturbative regime (and can
be set to zero independently without inconsistency). 
Hence we consider a PDF of the form
\begin{equation} \label{like}
p(x)= \frac{\rmn{e}^{-x^2/(2\sigma_0^2)}}{\sqrt{2 \pi}
\sigma_{0}} 
\left[ \alpha_{0} + \frac{\alpha_{3}}{\sqrt{48}} 
H_{3} \left(\frac{x}{\sqrt{2} \sigma_{0}}\right) \right]^{2}, 
\end{equation}
with $\alpha_0= \sqrt{1- \alpha_3^{2}}$.  It is straightforward to
show that the first, second and third moments of our PDF 
are related to $\alpha_3$ and $\sigma_{0}$ by \citep{contaldi}
\begin{eqnarray}
\mu_1&=&0  \nonumber\\
\mu_2&=&\sigma_0^{2}\left( 1+6\alpha_3^{2} \right) \nonumber\\
\mu_3&=&\left( 2\sigma_0^{2} \right)^\frac{3}{2} 
	\sqrt{3 \left[ \alpha_3^{2} \left(1-\alpha_3^{2} \right) \right]}.
\end{eqnarray}
The PDF therefore has zero mean and a fixed variance and skewness.  In
the simulations discussed in the main text, we choose $\alpha_{3}=0.2$
and $\sigma_{0}=1$. This resulting PDF is plotted in
Fig.~\ref{fig:sho}.

We note that the space of possible PDFs is constrained as a result of
restricting the set of coefficients $\alpha_{n}$ to two non-zero
values.  This implies that we cannot generate distributions with
arbitrarily large relative skewness. Indeed, $\mu_3/\mu_2^{3/2}$ is bounded
above by 0.74, and takes this maximum value for $\alpha_3^2 =(7-\sqrt{43})/6=
0.27^2$. However, in general our method can generate higher values of the
relative skewness (since it can generate any distribution) but for that
purpose one needs more non-zero coefficients $\alpha_n$ \citep{contaldi}.


\section{Some useful integrals}
\label{sec:integrals}

We give here useful results concerning integrals involving products of
Legendre polynomials:
\begin{equation}
I_{\ell_1\cdots \ell_n}(\mathbfit{x}_1,\ldots, \mathbfit{x}_n)
\equiv 
\int_{4\pi}\!\!
P_{\ell_1}(\mathbfit{x}_1\cdot\mathbfit{x})\cdots
P_{\ell_n}(\mathbfit{x}_n\cdot\mathbfit{x})\,\rmn{d}\Omega.
\end{equation}
Using the addition theorem (equation~\ref{eq:addition}), this reduces to
evaluating integrals of products of spherical harmonics,
\begin{equation}
J_{\ell_1 m_1,\ldots, \ell_n m_n}
\equiv \int_{4\pi}\!\! Y_{\ell_1 m_1}(\mathbfit{x}) \cdots
Y_{\ell_n m_n}(\mathbfit{x})\, \rmn{d}\Omega
\end{equation}
since
\begin{equation}
I_{\ell_1, \cdots, \ell_n}(\mathbfit{x}_1,\ldots,\mathbfit{x}_n)
= \frac{4 \pi}{2 \ell_1 +1} \cdots \frac{4 \pi}{2 \ell_n +1}
\sum_{m_1 \cdots m_n} Y^*_{\ell_1 m_1}(\mathbfit{x}_1) \cdots
Y^*_{\ell_n m_n}(\mathbfit{x}_n) J_{\ell_1 m_1,\ldots, \ell_n m_n}.
\end{equation}

First, consider the case $n=2$. Using the orthonormality of the spherical
harmonics, and the relation $Y_{\ell m}^*(\mathbfit{x}) =
(-1)^m Y_{\ell -m}(\mathbfit{x})$, to evaluate $J_{\ell_1 m_1,\ell_2 m_2}=
(-1)^{m_1} \delta_{\ell_1 \ell_2} \delta_{m_1\, -m_2}$, and then applying the
addition theorem we find the well-known result
\be
I_{\ell_1 \ell_2}(\mathbfit{x}_1,\mathbfit{x}_2) = \frac{4\pi}{2\ell_1+1}
P_{\ell_1}(\mathbfit{x}_1 \cdot \mathbfit{x}_2) \delta_{\ell_1 \ell_2}.
\label{eqn:neq2}
\ee
For integrals involving products of three or more spherical harmonics,
the general strategy is to combine pairs of harmonics using the
Clebsch-Gordan series (e.g.~\citealt{Varshalovich:1988,Edmonds:1974})
\begin{equation}
Y_{\ell_1 m_1}(\mathbfit{x}) Y_{\ell_2 m_2}(\mathbfit{x}) 
= \sum_{\ell m} \sqrt{\frac{(2 \ell_1 +1)(2 \ell_2 +1)(2 \ell +1)}{4 \pi}}
\left( 
\begin{array}{ccc}\ell_{1} & \ell_{2} & \ell\\
	          m_{1} & m_{2} & m
\end{array}
\right)
\left( 
\begin{array}{ccc}\ell_{1} & \ell_{2} & \ell\\
	          0 & 0 & 0
\end{array}
\right) 
Y^*_{\ell m}(\mathbfit{x}),
\label{eq:CG}
\end{equation}
until we have only a single pair left which can then be integrated
trivially using orthonormality.
We illustrate this procedure for the case of $n=3$ and $n=4$ since these
are needed for the calculation of the bispectrum and trispectrum. For
$n=3$, from equation~(\ref{eq:CG}) we find immediately that
\begin{equation}
J_{\ell_1 m_1,\ldots,\ell_3 m_3} = 
\sqrt{\frac{(2 \ell_1 +1)(2 \ell_2 +1)(2 \ell_3 +1)}{4 \pi}}
\left( 
\begin{array}{ccc}\ell_{1} & \ell_{2} & \ell_{3}\\
	          0 & 0 & 0
\end{array}
\right)
\left( 
\begin{array}{ccc}\ell_{1} & \ell_{2} & \ell_{3}\\
	          m_{1} & m_{2} & m_{3}
\end{array}
\right) ,
\end{equation}
and therefore
\begin{equation}
I_{\ell_1 \ell_2 \ell_3}(\mathbfit{x}_1, \mathbfit{x}_2, \mathbfit{x}_3)
= (4 \pi)^{2} \sqrt{\frac{4 \pi}{(2 \ell_1 +1)(2 \ell_2 +1) (2 \ell_3+1) }} 
\left( 
\begin{array}{ccc}\ell_{1} & \ell_{2} & \ell_{3}\\
	          0 & 0 & 0
\end{array}
\right)
\sum_{m_1 m_2 m_3}
\left( 
\begin{array}{ccc}\ell_{1} & \ell_{2} & \ell_{3}\\
	          m_{1} & m_{2} & m_{3}
\end{array}
\right) 
Y^*_{\ell_1 m_1}(\mathbfit{x}_1) 
Y^*_{\ell_2 m_2}(\mathbfit{x}_2) 
Y^*_{\ell_3 m_3}(\mathbfit{x}_3). 
\label{eqn:n3}
\end{equation}
The final term in this equation (the summation over $m_1$, $m_2$ and $m_3$)
ensures that $I_{\ell_1 \ell_2 \ell_3}$ is invariant under rigid rotations
of its vector arguments $\mathbfit{x}_1$, $\mathbfit{x}_2$ and $\mathbfit{x}_3$.
As expected, the summation can be expressed in terms of the tripolar
spherical harmonics with zero total angular momentum~\citep{Varshalovich:1988}.

Consider now the case $n=4$. There is now some freedom in the choice of
spherical harmonics to combine. If we couple $Y_{\ell_1 m_1}$ with $Y_{\ell_2
m_2}$ and $Y_{\ell_3 m_3}$ with $Y_{\ell_4 m_4}$, we find
\begin{eqnarray}
J_{\ell_1 m_1,\ldots, \ell_4 m_4} &=& 
\sum_{ \ell m} (-1)^{m} \sqrt{\frac{(2 \ell_1 +1)(2 \ell_2 +1) (2 \ell +1)}{4 \pi}} \sqrt{\frac{(2 \ell_3 +1)(2 \ell_4 +1)(2 \ell +1)}{4 \pi}}\\
& & \qquad\times 
\left( 
\begin{array}{ccc}\ell_{1} & \ell_{2} & \ell\\
	          m_{1} & m_{2} &  m
\end{array}
\right)
\left( 
\begin{array}{ccc}\ell_{3} & \ell_{4} & \ell\\
	          m_{3} & m_{4} &  -m
\end{array}
\right)
\left( 
\begin{array}{ccc}\ell_{1} & \ell_{2} & \ell\\
	           0 & 0 &  0
\end{array}
\right)
\left( 
\begin{array}{ccc}\ell_{3} & \ell_{4} & \ell\\
	           0 & 0 &  0
\end{array}
\right).
\end{eqnarray}
The expression on the right is not manifestly symmetric with
respect to interchange of e.g.\ $(\ell_1 m_1)$ and $(\ell_3 m_3)$ since the
latter involves a different coupling scheme.
However, the symmetry is easily verified by switching between the
two schemes with the 6$j$ coefficients~\citep{Varshalovich:1988}.
Finally, we find that
\begin{eqnarray}
I_{\ell_1\cdots \ell_4}(\mathbfit{x}_1,\ldots,\mathbfit{x}_4)
&=& (4\pi)^{3/2}\sqrt{\frac{(2\ell_1+1)\cdots(2\ell_4+1)}{4\pi}}
\sum_l \frac{2l+1}{4\pi}
\left( 
\begin{array}{ccc}\ell_{1} & \ell_{2} & \ell\\
	           0 & 0 &  0
\end{array}
\right)
\left( 
\begin{array}{ccc}\ell_{3} & \ell_{4} & \ell\\
	           0 & 0 &  0
\end{array}
\right) \nonumber \\
&&\mbox{} \times
\sum_{m m_1 \cdots m_4}
(-1)^m
\left( 
\begin{array}{ccc}\ell_{1} & \ell_{2} & \ell\\
	          m_{1} & m_{2} &  m
\end{array}
\right)
\left( 
\begin{array}{ccc}\ell_{3} & \ell_{4} & \ell\\
	          m_{3} & m_{4} &  -m
\end{array}
\right)
Y_{\ell_1 m_1}^*(\mathbfit{x}_1)Y_{\ell_2 m_2}^*(\mathbfit{x}_2)
Y_{\ell_3 m_3}^*(\mathbfit{x}_3)Y_{\ell_4 m_4}^*(\mathbfit{x}_4).
\end{eqnarray}
It is straightforward to verify that the last term on the right (the
summation over $m$, $m_1,\ldots,m_4$) is invariant under rigid rotations of
$\mathbfit{x}_1,\ldots,\mathbfit{x}_4$.


\section{Flat-Sky Approximation}
\label{sec:flat_sky}

For analysis over a small patch of the sky we can use the flat-sky
approximation and replace spherical transforms by Fourier transforms.
Our starting point is again a pixelised map of non-Gaussian white noise,
with each pixel value drawn from the non-Gaussian PDF $f_S(s)$.
Approximating the Fourier transform $a(\bell)$ by a discrete Fourier
transform we have
\begin{equation}
a(\bell) = \int\frac{\rmn{d}^2 \mathbfit{x}}{2\pi}\,
S(\mathbfit{x}) \rmn{e}^{-\rmn{i}\bell
\cdot \mathbfit{x}} \approx \frac{\Omega_{\rmn{pix}}}{2\pi} \sum_p
s_p \rmn{e}^{-\rmn{i} \bell\cdot \mathbfit{x}_p}, 
\end{equation}
where $\Omega_{\rmn{pix}}$ is the pixel area. We evaluate $a(\bell)$
on a regular grid in Fourier space with a Fast Fourier Transform. For a
square patch of sky with $N_{\rmn{pix}}$ pixels, the cell size in
Fourier space is $(2\pi)^2/(N_{\rmn{pix}} \Omega_{\rmn{pix}})$.
The second-order correlator of the discrete $a(\bell)$ evaluates
to
\begin{equation}
\langle a(\bell) a^*(\bell') \rangle = \left(\frac{\Omega_{\rmn{pix}}}
{2\pi}\right)^2 N_{\rmn{pix}} \mu_2 \delta_{\bell \bell'},
\label{eq:B2}
\end{equation}
where $\mu_2$ is the variance of the zero-mean $f_S(s)$.
In the continuum limit, equation~(\ref{eq:B2}) becomes
\begin{equation}
\langle a(\bell) a^*(\bell') \rangle = \Omega_{\rmn{pix}} \mu_2
\delta(\bell-\bell'),
\end{equation}
where we have used
\begin{equation}
\delta_{\bell\bell'} = \frac{1}{N_{\rmn{pix}}} \sum_p
\rmn{e}^{\rmn{i} (\bell-\bell')\cdot\mathbfit{x}_p} \rightarrow
\frac{1}{N_{\rmn{pix}}\Omega_{\rmn{pix}}} \int \rmn{d}^2\mathbfit{x}\,
\rmn{e}^{\rmn{i}  (\bell-\bell')\cdot\mathbfit{x}} = \frac{(2\pi)^2}
{N_{\rmn{pix}}\Omega_{\rmn{pix}}} \delta(\bell-\bell').
\end{equation}
We scale the $a(\bell)$ defining $\bar{a}(\bell) \equiv
\sqrt{C_\ell/(\Omega_{\rmn{pix}} \mu_2)} a(\bell)$ (as for the full-sky
case described in the main text) such that the
$\bar{a}(\bell)$ have the required power spectrum:
\begin{equation}
\langle \bar{a}(\bell) \bar{a}^*(\bell') \rangle = C_\ell \delta(\bell
- \bell').
\end{equation}
Finally, we inverse Fourier transform to obtain our non-Gaussian map,
$T(\mathbfit{x})$, with the prescribed two-point statistics:
\begin{equation}
T(\mathbfit{x}_p) = \frac{2\pi}{N_{\rmn{pix}}\Omega_{\rmn{pix}}} \sum_{\bell}
\bar{a}(\bell) \rmn{e}^{\rmn{i} \bell \cdot \mathbfit{x}_p} \approx
\int \frac{\rmn{d}^2 \bell}{2\pi} \bar{a}(\bell)
\rmn{e}^{\rmn{i} \bell \cdot \mathbfit{x}_p}.
\end{equation}

As in the full-sky case, we can express the pixel values, $t_p$,
in the final map as linear combinations of those in the original map,
$s_p$:
\begin{equation}
t_p = \sum_{p'} W_{pp'} s_{p'},
\end{equation}
where in the continuum approximation
\begin{eqnarray}
W_{pp'} &=& \frac{1}{2\pi}\sqrt{\frac{\Omega_{\rmn{pix}}}{\mu_2}}
\int \frac{\rmn{d}^2 \bell}{2\pi} \, \sqrt{C_\ell} \rmn{e}^{\rmn{i}\bell\cdot (\mathbfit{x}_p
- \mathbfit{x}_{p'})} \\
&=& \sqrt{\frac{\Omega_{\rmn{pix}}}{\mu_2}} \int \frac{\ell d \ell}{2\pi} \sqrt{C_\ell}
J_0(\ell|\mathbfit{x}_p - \mathbfit{x}_{p'}|), \nonumber
\end{eqnarray}
with $J_0(z)$ the Bessel function of order zero. Using the asymptotic
result $J_0(\ell\theta) \approx P_\ell(\cos\theta)$, it is straightforward to see
that $W_{pp'}$ obtained here in the flat-sky limit is equivalent to the
full-sky expression (equation~\ref{eqn:wdef}).
Forming the connected $n$-point function,
as in equation~(\ref{eq:correlators}), we find
\begin{eqnarray}
\langle t_{p_1} \cdots t_{p_n} \rangle_c &=&
\frac{\kappa_n}{\Omega_{\rmn{pix}}}
\left(\frac{\Omega_{\rmn{pix}}}{\mu_2}\right)^{n/2} \int \frac{\rmn{d}^2 \bell_1}
{(2\pi)^2} \cdots \frac{\rmn{d}^2 \bell_n}{(2\pi)^2} \, \sqrt{C_{\ell_1}
\cdots C_{\ell_n}} \int \rmn{d}^2 \mathbfit{x}\, \rmn{e}^{\rmn{i}\bell_1 \cdot (\mathbfit{x}_{p_1}-
\mathbfit{x})} \cdots \rmn{e}^{\rmn{i}\bell_n \cdot (\mathbfit{x}_{p_n}-\mathbfit{x})} \nonumber
\\
&=& (2\pi)^2 \frac{\kappa_n}{\Omega_{\rmn{pix}}}
\left(\frac{\Omega_{\rmn{pix}}}{\mu_2}\right)^{n/2} \int \frac{\rmn{d}^2 \bell_1}
{(2\pi)^2} \cdots \frac{\rmn{d}^2 \bell_n}{(2\pi)^2} \, \sqrt{C_{\ell_1}
\cdots C_{\ell_n}}\delta(\bell_1 + \cdots + \bell_n)
\rmn{e}^{\rmn{i}(\bell_1 \cdot \mathbfit{x}_{p_1} + \cdots +
\bell_n \cdot \mathbfit{x}_{p_n})}.
\end{eqnarray}
For example, for $n=2$ we have
\begin{equation}
\langle t_{p_1} t_{p_2} \rangle_c = \int \frac{\rmn{d}^2 \bell}{(2\pi)^2}
C_\ell \rmn{e}^{\rmn{i}\bell\cdot(\mathbfit{x}_{p_1} - \mathbfit{x}_{p_2})}
\approx \int \frac{\ell d \ell}{2\pi} C_\ell J_0(\ell|\mathbfit{x}_{p_1} - \mathbfit{x}_{p_2}|).
\end{equation}
In the asymptotic limit this result reduces to the full-sky expression
(equation~\ref{eq:2point}).

Finally, we consider the polyspectra of the processed non-Gaussian maps.
Expressing the Fourier transform of the final map, $\bar{a}(\bell)$,
in terms of the original map, i.e.\
\begin{equation}
\bar{a}(\bell) = \frac{1}{2\pi}\sqrt{\frac{C_\ell \Omega_{\rmn{pix}}}{\mu_2}}
\sum_p s_p \rmn{e}^{-\rmn{i}\bell\cdot \mathbfit{x}_p},
\end{equation}
we have
\begin{eqnarray}
\langle \bar{a}(\bell_1) \cdots \bar{a}(\bell_n) \rangle_c &=&
\frac{\kappa_n}{(2\pi)^n} \left(\frac{\Omega_{\rmn{pix}}}{\mu_2}\right)^{n/2}
\sqrt{C_{\ell_1} \cdots C_{\ell_n}} \sum_p \rmn{e}^{-\rmn{i}(\bell_1 + \cdots + \bell_n)
\cdot \mathbfit{x}_p} \nonumber \\
&\approx& \frac{\kappa_n}{\Omega_{\rmn{pix}}}
\left(\frac{\Omega_{\rmn{pix}}}{\mu_2}\right)^{n/2} (2\pi)^{2-n}
\sqrt{C_{\ell_1} \cdots C_{\ell_n}} \delta(\bell_1 + \cdots + \bell_n),
\end{eqnarray}
where in the last line we have taken the continuum approximation.
This form for the correlator is clearly consistent with rotational,
translational and parity invariance. 

As in the full-sky case, we have produced simulated non-Gaussian maps
and calculated their power spectra and bispectra; the latter 
were estimated using the code described in~\cite{smith04}. We find
that the flat-sky simulations behave as expected with no
discernible bias.

\label{lastpage}
\bsp
\end{document}